\documentstyle[12pt]{article}

\begin{document}

\newcommand{\nc}[2]{\newcommand{#1}{#2}}
\newcommand{\ncx}[3]{\newcommand{#1}[#2]{#3}}
\ncx{\pr}{1}{#1^{\prime}}
\nc{\nl}{\newline}
\nc{\np}{\newpage}
\nc{\nit}{\noindent}
\nc{\be}{\begin{equation}}
\nc{\ee}{\end{equation}}
\nc{\ba}{\begin{array}}
\nc{\ea}{\end{array}}
\nc{\bea}{\begin{eqnarray}}
\nc{\eea}{\end{eqnarray}}
\nc{\nb}{\nonumber}
\nc{\dsp}{\displaystyle}
\nc{\bit}{\bibitem}
\nc{\ct}{\cite}
\ncx{\dd}{2}{\frac{\partial #1}{\partial #2}}
\nc{\pl}{\partial}
\nc{\dg}{\dagger}
\nc{\wig}{\wedge}
\nc{\cL}{{\cal L}}
\nc{\cD}{{\cal D}}
\nc{\cF}{{\cal F}}
\nc{\cG}{{\cal G}}
\nc{\cJ}{{\cal J}}
\nc{\cQ}{{\cal Q}}
\nc{\tB}{\tilde{B}}
\nc{\tD}{\tilde{D}}
\nc{\tH}{\tilde{H}}
\nc{\tQ}{\tilde{Q}}
\nc{\tR}{\tilde{R}}
\nc{\tZ}{\tilde{Z}}
\nc{\tg}{\tilde{g}}
\nc{\tog}{\tilde{\og}}
\nc{\tGam}{\tilde{\Gam}}
\nc{\tPi}{\tilde{\Pi}}
\nc{\tcD}{\tilde{\cD}}
\nc{\tcQ}{\tilde{\cQ}}
\nc{\ag}{\alpha}
\nc{\bg}{\beta}
\nc{\gam}{\gamma}
\nc{\Gam}{\Gamma}
\nc{\bgm}{\bar{\gam}}
\nc{\del}{\delta}
\nc{\Del}{\Delta}
\nc{\eps}{\epsilon}
\nc{\ve}{\varepsilon}
\nc{\zg}{\zeta}
\nc{\th}{\theta}
\nc{\vt}{\vartheta}
\nc{\Th}{\Theta}
\nc{\kg}{\kappa}
\nc{\lb}{\lambda}
\nc{\Lb}{\Lambda}
\nc{\ps}{\psi}
\nc{\Ps}{\Psi}
\nc{\sg}{\sigma}
\nc{\spr}{\pr{\sg}}
\nc{\Sg}{\Sigma}
\nc{\rg}{\rho}
\nc{\fg}{\phi}
\nc{\Fg}{\Phi}
\nc{\vf}{\varphi}
\nc{\og}{\omega}
\nc{\Og}{\Omega}
\nc{\Kq}{\mbox{$K(\vec{q},t|\pr{\vec{q}\,},\pr{t})$ }}
\nc{\Kp}{\mbox{$K(\vec{q},t|\pr{\vec{p}\,},\pr{t})$ }}
\nc{\vq}{\mbox{$\vec{q}$}}
\nc{\qp}{\mbox{$\pr{\vec{q}\,}$}}
\nc{\vp}{\mbox{$\vec{p}$}}
\nc{\va}{\mbox{$\vec{a}$}}
\nc{\vb}{\mbox{$\vec{b}$}}
\nc{\Ztwo}{\mbox{\sf Z}_{2}}
\nc{\Tr}{\mbox{Tr}}
\nc{\lh}{\left(}
\nc{\rh}{\right)}
\nc{\ld}{\left.}
\nc{\rd}{\right.}
\nc{\nil}{\emptyset}
\nc{\bor}{\overline}
\nc{\ha}{\hat{a}}
\nc{\da}{\hat{a}^{\dg}}
\nc{\hb}{\hat{b}}
\nc{\db}{\hat{b}^{\dg}}
\nc{\hN}{\hat{N}}
\ncx{\abs}{1}{\left| #1 \right|}

\pagestyle{empty}

\begin{flushright}
NIKHEF-H/93-25
\end{flushright}

\begin{center}
{\LARGE {\bf New supersymmetries for spinning}} \\
\vspace{3ex}
{\LARGE {\bf particles and black holes}}\footnote{Lecture presented at the
workshop on Contraint Theory and Quantization Methods, Montepulciano (It.),
June 28 -- July 1, 1993} \\

\vspace{5ex}

{\large J.W.\ van Holten} \\
NIKHEF-H, Amsterdam NL \\

\vspace{7ex}

{\bf Abstract} \\
\end{center}

\nit
{\small
The usual extensions of supersymmetry require the existence of a complex
structure and the formulation of the theory on K\"{a}hler manifolds. It is
shown, that by relaxing the constraints on the algebra of supercharges we can
get new supersymmetries whenever a manifold possesses a structure admitting the
existence of a Killing-Yano tensor field. Examples of such manifolds are the
Kerr-Newman space-times describing spinning black holes in four dimensions. }

\np

\pagestyle{plain}
\pagenumbering{arabic}

\section{Introduction}

Spinning black holes in four-dimensional space-time, electrically neutral or
charged, are described by the Kerr and Kerr-Newman solutions of the pure
Einstein and coupled Einstein-Maxwell equations, respectively. An explicit form
of these static solutions, of mass $M$, charge $e$ and angular momentum $ J = M
a$, is represented by the line-element

\be
\ba{lll}
ds^2 & = &\dsp{ - \frac{\Del}{\rg^2}\, \lh dt - a \sin^2 \th d \fg \rh^2\, +\,
             \frac{\sin^2 \th}{\rg^2}\, \lh (r^2 + a^2)d\fg - a dt \rh^2 }\\
  &  & \\
  &  & \dsp{ +\, \frac{\rg^2}{\Del}\, dr^2\, +\, \rg^2 d\th^2, }
\ea
\label{1.1}
\ee

\nit
where we have used the abbreviations

\be
\rg^2\, =\, r^2\, +\, a^2\, \cos^2 \th, \hspace{3em}
\Del\, =\, r^2\, +\, a^2\, - 2 Mr\, +\, e^2\, >\, 0.
\label{1.2}
\ee

\nit
In case of non-vanishing charge $e$ the corresponding electro-magnetic field
is described by the Maxwell 2-form

\be
\ba{lll}
F & = & \dsp{ \frac{e}{\rg^4}\, \lh r^2 - a^2 \cos^2 \th \rh\, dr\, \wig\,
    \lh dt - a \sin^2 \th d\fg \rh }\\
  & & \\
  & & \dsp{ +\, 2\, \frac{ear\cos \th \sin \th}{\rg^4}\, d\th\, \wig\,
            \lh (r^2 + a^2) d\fg - a dt \rh. }
\ea
\label{1.3}
\ee

\nit
It was discovered by Carter \ct{C} that the geodesic equations for the
four-dimensional space-time geometry described by this metric are completely
integrable in the sense of Liouville. Closely connected to this is the
existence
of a constant of motion $Z$ for a point-particle moving in Kerr-Newman
space-time, of the form

\be
Z\, =\, \frac{1}{2}\, K_{\mu\nu}(x) \dot{x}^{\mu} \dot{x}^{\nu}, \hspace{3em}
\dot{Z}\, =\, 0.
\label{1.4}
\ee

\nit
The overdot here denotes proper-time differentiation, and the symmetric tensor
field $K_{\mu\nu}$ is a Killing-tensor, implying that its completely
symmetrized
covariant derivative vanishes:

\be
K_{\lh\mu\nu;\lb\rh}\, =\, 0.
\label{1.5}
\ee

\nit
In the co-ordinate system (\ref{1.1}) the explicit form of $Z$ is

\be
\ba{lll}
Z & = & \dsp{ \frac{\Del \cos^2 \th}{\rg^2}\, \lh \dot{t} - a \sin^2 \th
              \dot{\fg} \rh^2\, +\, \frac{r^2 \sin^2 \th}{\rg^2 a^2}\,
              \lh (r^2 + a^2) \dot{\fg} - a\dot{t} \rh^2 }\\
  &  & \\
  &  & \dsp{ -\, \frac{\rg^2 \cos^2 \th}{\Del}\, \dot{r}^2\, +\,
             \frac{\rg^2}{r^2}{a^2}\, \dot{\th}^2.}
\ea
\label{1.6}
\ee

\nit
Note that the symmetric quadratic differential $K_{\mu\nu} d x^{\mu} d x^{\nu}
=
2 Z d \tau^2$ is {\em different} from the line-element $ds^2$; the constant of
motion corresponding to the latter is the world-line hamiltonian $H$.

In correpondence with these classical results is the observation by the
authors of \ct{C2,Ch} that the Klein-Gordon and Dirac equations in the curved
space-time (\ref{1.1}) are separable. For the Klein-Gordon equation this
follows
from the conservation of the operator corresponding to $Z$, eq.(\ref{1.4}). In
the case of the Dirac equation it was established in \ct{P,F} that there exists
a square root of the Killing tensor $K_{\mu\nu}$:

\be
K_{\mu\nu}\, =\, \eta_{ab} f_{\mu}^{\:\:a} f_{\nu}^{\:\:b},
\label{1.7}
\ee

\nit
which is anti-symmetric after contraction with the vierbein:

\be
f_{\mu\nu}\, \equiv f_{\mu}^{\:\:a} e_{\nu a}\, =\, -f_{\nu\mu},
\label{1.8}
\ee

\nit
and such that one can construct from it a linear differential operator which
anti-commutes with the Dirac operator. This result depends crucially on the
complete anti-symmetry of the covariant derivative of $f_{\mu\nu}$:

\be
H_{\mu\nu\lb}\, =\, f_{[\mu\nu ; \lb]}\, =\, f_{\mu\nu ; \lb}.
\label{1.9}
\ee

\nit
Anti-symmetric tensors of this kind are known as Killing-Yano tensors; their
appearance is not restricted to the spinning black-hole solutions of general
relativity. In the following we will see that they are closely connected to
the existence of certain new types of supersymmetries, of which the spinning
black holes provide only one example, though an interesting one. A systematic
analysis of these supersymmetries is presented in \ct{GRH}.

\section{Spinning particles as supersymmetric {$\sg$}-models}

In \ct{RJW1,RJW2} it was shown how Killing vectors and tensors related to
symmetries of manifolds, and their fermionic extensions for spinning manifolds,
can be obtained from the geodesic equations of motion for point particles
and spinning particles in curved background space-times. The Lie-algebra of
the Killing vector and tensor fields is reflected in the algebra of
Poisson-Dirac brackets of the constants of motion of these physical systems.
The starting point is the lagrangian for a spinning particle in a gravitational
field

\be
L\, =\, \frac{1}{2}\, g_{\mu\nu} \dot{x}^{\mu} \dot{x}^{\nu}\, +\,
        \frac{i}{2}\, \eta_{ab} \ps^{a} \frac{D \ps^{b}}{D\tau}.
\label{2.1}
\ee

\nit
The action is invariant under the supersymmetry

\be
\del x^{\mu}\, =\, - i \eps \ps^{\mu}, \hspace{3em}
\del \ps^{\mu}\, =\, \dot{x}^{\mu} \eps,
\label{2.2}
\ee

\nit
where we use the notation $\ps^{\mu}(x,\ps)\, =\, e^{\mu}_{\:\:a} \ps^{a}$,
and $x^{\mu}(\tau)$ represents the position variables, whilst the
Grassmann-odd variables $\ps^{a}(\tau)$, transforming as a local Lorentz
vector,
describe the spin.

The equations of motion obtained from the variation of the lagrangian
(\ref{2.1}) are

\begin{eqnarray}
\dsp{ \frac{D^2 x^{\mu}}{D \tau^2} } & = & \dsp{ \ddot{x}^{\mu}\, -\,
      \Gam_{\lb\nu}^{\:\:\:\:\:\mu}\, \dot{x}^{\lb} \dot{x}^{\nu}\, =\,
      -\frac{i}{2}\, \ps^a \ps^b\, R_{ab\:\:\nu}^{\:\:\:\:\:\mu}
\dot{x}^{\nu},}
\label{2.3}\\
  &  &  \nonumber \\
\dsp{ \frac{D \ps^a }{D\tau} } & = & \dsp{ \dot{\ps}^a\, -\, \dot{x}^{\mu}
      \og_{\mu \:\:b}^{\:\:a}\, \ps^b\, =\, 0, }
\label{2.4}
\end{eqnarray}

\nit
with
$\og_{\mu\:\:b}^{\:\:a}$ the spin connection and
$R_{ab\:\:\nu}^{\:\:\:\:\:\mu}$
the Riemann tensor. In terms of the spin tensor $S^{ab} = - i \ps^a \ps^b$ (an
anti-symmetric local Lorentz tensor) this becomes

\be
\frac{D^2 x^{\mu}}{D\tau^2}\, =\, \frac{1}{2}\, S^{ab}
                                  R_{ab\:\:\nu}^{\:\:\:\:\:\mu} \dot{x}^{\nu},
\hspace{2em} \frac{D S^{ab} }{D\tau}\, =\, 0.
\label{2.5}
\ee

The above theory can be recast in canonical form; we introduce the momentum

\be
p_{\mu}\, =\, g_{\mu\nu} \dot{x}^{\nu}\, +\, \og_{\mu},
\label{2.6}
\ee

\nit
and hamiltonian

\be
H\, =\, \frac{1}{2}\, g^{\mu\nu}\, (p_{\mu} - \og_{\mu})(p_{\nu} - \og_{\nu}),
\label{2.7}
\ee

\nit
in which

\be
\og_{\mu}\, =\, \frac{1}{2}\, \og_{\mu ab} S^{ab}
\label{2.8}
\ee

\nit
is the Grassmann-valued spin connection. In terms of the Poisson-Dirac brackets

\be
\left\{ F, G \right\}\, =\, \dd{F}{x^{\mu}} \dd{G}{p_{\mu}}\, - \dd{F}{p_{\mu}}
        \dd{G}{x^{\mu}}\, +\, i (-1)^{a_F}\, \dd{F}{\ps^a} \dd{G}{\ps_a},
\label{2.9}
\ee

\nit
with $a_F$ the Grassmann parity of $F$, the equations of motion for an
arbitrary
dynamical quantity $F$ can now be written as

\be
\dot{F}\, =\, \left\{ F, H \right\}.
\label{2.10}
\ee

\nit
With this defintion of the brackets it also follows, that the spin tensor
satisfies the Lorentz algebra

\be
\left\{ S^{ab}, S^{cd} \right\}\, =\, \eta^{ad} S^{bc}\, +\, \eta^{bc} S^{ad}\,
                                  -\, \eta^{ac} S^{bd}\, -\, \eta^{bd} S^{ac}.
\label{2.11}
\ee

\nit
This confirms the interpretation of $S^{ab}$ as the spin: it generates the
`internal' part of the Lorentz transformations on the phase space spanned by
$(x^{\mu}, p_{\mu}, \ps^a)$.

\section{Covariant phase space}

Results and calculations for the spinning particle theory can be simplified
by introducing  a covariant phase space formulation, obtained by changing to
the covariant momentum

\be
\Pi_{\mu}\, =\, p_{\mu}\, -\, \og_{\mu}\, =\, g_{\mu\nu} \dot{x}^{\nu}.
\label{3.1}
\ee

\nit
In terms of this variable the hamiltonian reads

\be
H\, =\, \frac{1}{2}\, g^{\mu\nu} \Pi_{\mu} \Pi_{\nu},
\label{3.2}
\ee

\nit
whilst the Poisson-Dirac brackets take the form

\be
\left\{ F, G \right\}\, =\, \cD_{\mu} F \dd{G}{\Pi_{\mu}}\, -\,
        \dd{F}{\Pi_{\mu}} \cD_{\mu} G\, +\, R_{\mu\nu} \dd{F}{\Pi_{\mu}}
        \dd{G}{\Pi_{\nu}}\, +\, i (-1)^{a_F}\, \dd{F}{\ps^a} \dd{G}{\ps_a}.
\label{3.3}
\ee

\nit
Here

\be
\cD_{\mu} F\, =\, \partial_{\mu} F\, +\, \Gam_{\mu\nu}^{\:\:\:\:\:\lb}
\Pi_{\lb}
                  \dd{F}{\Pi_{\nu}}\, +\, \og_{\mu ab} \ps^{b} \dd{F}{\ps_a}.
\label{3.4}
\ee

\nit
For scalar observables $F, G$ the covariant brackets always automatically
give covariant results. For the covariant momenta we obtain the classical
version of the Ricci identity:

\be
\left\{ \Pi_{\mu}, \Pi_{\nu} \right\}\, =\, R_{\mu\nu}\, =\, \frac{1}{2}\,
     S^{ab} R_{ab\mu\nu}.
\label{3.5}
\ee

\nit
Since in this formulation it is advantageous to work only with scalar
quantities, we observe that these can always be obtained from any tensorial
expression by saturating the indices with either covariant momenta $\Pi_{\mu}$
or spin variables $\ps^a$, depending on whether one needs to symmetrize or
anti-symmetrize. Thus all quantities we consider are of the form

\be
F(x,\Pi,\ps)\, =\,
   \sum_{m,n \geq 0}\, \frac{i^{\left[\frac{m}{2}\right]}}{m!n!}\,
   \ps^{a_1} ... \ps^{a_m}\, f^{\mu_1 ... \mu_n}_{a_1 ... a_m}(x)\,
   \Pi_{\mu_1} ... \Pi_{\mu_n}.
\label{3.6}
\ee

\nit
One may think of these scalar functions as generalized differential forms
on a graded phase space.

\section{Symmetries}

In the hamiltonian formulation, symmetry transformations are generated by
the constants of motion through the Poisson-Dirac brackets. In particular, the
supersymmetry transformations (\ref{2.2}) are obtained from the conserved
supercharge

\be
Q = \Pi \cdot \ps = e^{\mu}_{\:\:a}\, \Pi_{\mu} \ps^a, \hspace{3em}
          \dot{Q} = 0,
\label{4.1}
\ee

\nit
by taking the bracket

\be
\del F\, =\, i\eps \left\{ Q, F\right\}.
\label{4.2}
\ee

\nit
That $Q$ is conserved and the super-transformations (\ref{4.2}) represent a
symmetry follows from the bracket relations

\be
\left\{ Q, Q \right\}\, =\, - 2i H, \hspace{3em}
\left\{ Q, H \right\}\, =\, 0.
\label{4.3}
\ee

\nit
The second relation, which follows from the first by the Jacobi identity,
implies at the same time the conservation of $Q$ and the invariance of $H$
under the transformations (\ref{4.2}).

After the pattern established for the supercharge $Q$, we can now investigate
the full set of symmetries for a given space-time by solving the equation

\be
\left\{ J, H \right\}\, =\, \Pi^{\mu} \lh \cD_{\mu} J + R_{\mu\nu}\,
 \dd{J}{\Pi_{\nu}} \rh\, =\,  0,
\label{4.4}
\ee

\nit
which give all constants of motion $J(x,\Pi,\ps)$. This equation is the
generalization of the usual Killing equation to spinning space \ct{RJW1,RJW2}.
However, unlike the usual case, in which the solutions of the Killing
equation are single completely symmetric tensors, here the solutions consist
of linear combinations of symmetric tensors of different rank:

\be
J(x,\Pi,\ps)\, =\, \sum_{n \geq 0}\, \frac{1}{n!}\, J^{\mu_1 ... \mu_n}(x,\ps)
                   \Pi_{\mu_1} ... \Pi_{\mu_n},
\label{4.5}
\ee

\nit
subject to the conditions

\be
J_{\lh \mu_1 ... \mu_n ;\mu_{n+1}\rh}\, + \og_{\lh \mu_{n+1} \rd}^{\:\:\:ab}\,
       \ps_{b}\, \dd{J_{\ld \mu_1 ... \mu_n \rh}}{\ps^a}\, =\,
       - \frac{i}{2}\, \ps^a \ps^b R_{ab\:\lh\mu_{n+1}\rd}^{\:\:\:\:\:\nu}\,
       J_{\ld\mu_1 ... \mu_n \rh \nu}.
\label{4.6}
\ee

\nit
A sufficient, though not necessary, condition for a solution of the generalized
Killing equations is superinvariance of a dynamical variable:

\be
\left\{ J, Q \right\}\, =\, \ps \cdot \cD J\, +\, i \Pi \cdot \dd{J}{\ps}\,
                        =\, 0.
\label{4.7}
\ee

\nit
This equation may be considered as a kind of square root of the generalized
Killing equation. The new supersymmetries we present later satisfy this
superinvariance condition. A noteworthy exception is the supercharge $Q$
itself;
according to (\ref{4.3}) its bracket with $Q$ gives the hamiltonian, which
generates proper-time translations.

The solutions of the generalized Killing equation (\ref{4.4}) are of two
distinct types: {\em generic} ones, which exist for any spinning particle
model (\ref{2.1}), and {\em non-generic} ones, which depend on the specific
background space-time considered. To the first class belong supersymmetry
and proper-time translations, generated by the supercharge and hamiltonian,
respectively. In addition there also is a `chiral' symmetry, generated by the
conserved charge

\be
\Gam_*\, =\, - \frac{i^{\left[ \frac{d}{2} \right]}}{d!}\,
             \ve_{a_1 ... a_d}\, \ps^{a_1} ... \ps^{a_d},
\label{4.8}
\ee

\nit
and a dual supersymmetry generated by

\be
Q^* \, =\, i\, \left\{ Q, \Gam_* \right\}\, =\,
           - \frac{i^{\left[ \frac{d}{2} \right]}}{(d-1)!}\, e^{\mu a_1}\,
           \Pi_{\mu}\, \ve_{a_1 ... a_d}\, \ps^{a_2} ... \ps^{a_d}.
\label{4.9}
\ee

\nit
Note that $Q^*$ is Grassmann odd in even-dimensional space-times and Grassmann
even in odd-dimensional space-times. In the special case $d=2$ dual
supersymmetry is a real supersymmetry, in the sense that the bracket of $Q^*$
with itself closes on the hamiltonian. For all $d > 2$ this bracket vanishes
identically.

\section{New supersymmetries}

The existence of non-generic symmetries depends by definition on the
background space-time considered. We now ask, what are the necessary
conditions for the existence of new supersymmetries such that

\be
\del x^{\mu}\, =\, - i \eps\, f^{\mu}_{\:\:a}(x) \ps^a,
\label{5.1}
\ee

\nit
with $f^{\mu}_{\:\:a}$ some vector not equal to the vierbein $e^{\mu}_{\:\:a}$.
It is straightforward to establish that the solution to this problem is the
existence of a constant of motion

\be
Q_f\, =\, f^{\mu}_{\:\:a}\, \Pi_{\mu} \ps^a\, +\, \frac{i}{3!}\, c_{abc}\,
          \ps^a \ps^b \ps^c,
\label{5.2}
\ee

\nit
with the tensorial quantities $f^{\mu}_{\:\:a}$ and $c_{abc}$ subject to

\be
\ba{c}
D_{\mu} f^{\:\:a}_{\nu}\, +\, D_{\nu} f^{\:\:a}_{\mu}\, =\, 0, \\
  \\
D_{\mu} c_{abc}\, +\, R_{\mu\nu ab} f^{\nu}_{\:\:c}\, +\, R_{\mu\nu bc}
                  f^{\nu}_{\:\:a}\, +\, R_{\mu\nu ca} f^{\nu}_{\:\:b}\, =\, 0.
\ea
\label{5.3}
\ee

\nit
These conditions express the contents of the generalized Killing equation for
$Q_f$. The existence of a new supersymmetry of this kind then implies
automatically the existene of a new Grassmann-even constant of motion $Z$,
defined by the bracket of $Q_f$ with itself:

\be
\left\{ Q_f, Q_f \right\}\, =\,- 2i Z.
\label{5.4}
\ee

\nit
The explicit form of $Z$ is

\be
Z\, =\, \frac{1}{2}\, K^{\mu\nu}\, \Pi_{\mu} \Pi_{\nu}\, +\, I^{\mu}
\Pi_{\mu}\,
        +\, G,
\label{5.5}
\ee

\nit
with

\be
\ba{lll}
K^{\mu\nu} & = & K^{\nu\mu}\, =\, \eta_{ab} f^{\mu a} f^{\nu b}, \\
  & & \\
I^{\mu} & = & \dsp{ \frac{i}{2}\, \ps^a \ps^b \lh 2 f^{\nu}_{\:\:b} D_{\nu}
              f^{\mu}_{\:\:a}\, +\, f^{\mu c} c_{abc} \rh, }\\
  & & \\
G & = & \dsp{ - \frac{1}{4}\, \ps^a \ps^b \ps^c \ps^d \lh R_{\mu\nu ab}
        f^{\mu}_{\:\:c} f^{\nu}_{\:\:d}\, +\,
        \frac{1}{2}\, c_{ab}^{\:\:\:\:\:e} c_{cde} \rh. }
\ea
\label{5.6}
\ee

\nit
Since $Q_f$ satisfies the generalized Killing equation, and therefore

\be
\left\{ Q_f, H \right\}\, =\, 0,
\label{5.7}
\ee

\nit
the bracket relation (\ref{5.4}) in combination with the Jacobi identity
imply the conservation of $Z$:

\be
\left\{ Z, H \right\}\, =\, 0.
\label{5.8}
\ee

\nit
Therefore $Z$ is a solution of the generalized Killing equations as well,
and its components satisfy

\be
\ba{lll}
K_{ \lh \mu \nu ; \lb \rh } & = & 0, \\
  &  &  \\
D_{\lh \mu \rd} I_{\ld \nu \rh ab} & = & R_{ab\lb \lh \mu \rd}
                K_{\ld \nu \rh}^{\:\:\lb}, \\
  &  &  \\
D_{\mu} G_{abcd} & = & R_{\lb\mu\left[ab \rd}\, I^{\lb}_{\:\ld cd \right]},
\ea
\label{5.9}
\ee

\nit
where the square brackets in the last expression denote anti-symmetrization
over the latin indices enclosed.

\section{Killing-Yano tensors}

We now impose the requirement of the independence of the new supersymmetry
by requiring it to anti-commute with ordinary supersymmetry, or

\be
\left\{ Q, Q_f \right\}\, =\, 0.
\label{6.1}
\ee

\nit
As a direct consequence, we find that the second-rank tensor $f^{\mu\nu} =
f^{\mu a} e^{\nu}_{\:\:a}$ is anti-symmetric:

\be
f^{\mu\nu}\, +\, f^{\nu\mu}\, =\, 0.
\label{6.2}
\ee

\nit
In addition, the first Killing equation (\ref{5.3}) implies that

\be
f_{\mu\nu ;\lb} + f_{\lb\nu ;\mu}\, =\, 0.
\label{6.3}
\ee

\nit
Combining the two results yields the complete anti-symmetry of the covariant
derivative:

\be
\ba{lll}
f_{\mu\nu ;\lb} & = & \dsp{ \frac{1}{3}\, \lh f_{\mu\nu ;\lb} + f_{\nu\lb ;\mu}
                            + f_{\lb\mu ;\nu} \rh }\\
  &  &  \\
                & = & H_{\mu\nu\lb},
\ea
\label{6.4}
\ee

\nit
the field strength of the anti-symmetric tensor $f_{\mu\nu}$. It follows that
$f_{\mu\nu}$ is a Killing-Yano tensor.

Further differentiation of $H_{\mu\nu\lb}$ then leads to the result

\be
H_{\mu\nu\lb ;\kg}\, =\, \frac{1}{2}\, \lh R_{\nu\lb\mu}^{\:\:\:\:\:\:\:\sg}
   f_{\sg\kg} + R_{\lb\kg\mu}^{\:\:\:\:\:\:\:\sg} f_{\sg\nu} +
   R_{\kg\nu\mu}^{\:\:\:\:\:\:\:\sg} f_{\sg\lb} \rh.
\label{6.5}
\ee

\nit
Comparing with the second Killing equation (\ref{5.3}) we conclude, that
it is solved by taking

\be
c_{abc}\, =\, - \frac{1}{2}\, H_{abc},
\label{6.6}
\ee

\nit
the local Lorentz 3-form corresponding to the field strength tensor. Thus,
given
a Killing-Yano tensor $f_{\mu\nu}$, such an anti-symmetric 3rd rank tensor
always exists.

We therefore conclude that the existence of a Killing-Yano tensor is both a
necessary and a sufficient condition for the existence of a new supersymmetry
of the type (\ref{5.2}), obeying the condition (\ref{6.1}).

\section{Kerr-Newman geometry}

In the introduction we presented the uniformly rotating Kerr-Newman solution of
the Einstein equations, and mentioned the existence of a Kiling-Yano tensor in
this space-time; its explicit components are

\be
\ba{lll}
f_{\mu}^{\:\:0}\, d x^{\mu} & = & \dsp{ \frac{\rg}{\Del}\, \cos \th\, dr, }\\
  &  & \\
f_{\mu}^{\:\:1}\, d x^{\mu} & = & \dsp{ - \frac{\sqrt{\Del}}{\rg}\, \cos \th\,
                                  \lh dt - a \sin^2 \th\, d\fg \rh, }\\
  &  & \\
f_{\mu}^{\:\:2}\, d x^{\mu} & = & \dsp{ - \frac{r \sin \th}{a \rg}\,
                                  \lh (r^2 + a^2) d\fg - a dt \rh, }\\
  &  & \\
f_{\mu}^{\:\:3}\, d x^{\mu} & = & \dsp{ - \frac{r \rg}{a}\, d\th.}
\ea
\label{7.1}
\ee

\nit
The corresponding components of the Lorentz 3-form $c_{abc}$
obtained from the field strength are:

\be
\ba{lll}
c_{012} & = & \dsp{ \frac{2\sin\th}{\rg}, }\\
 & & \\
c_{123} & = & \dsp{ -\frac{2 \sqrt{\Del}}{a \rg}, }\\
 & & \\
c_{013} & = & c_{023}\, =\, 0.
\ea
\label{7.2}
\ee

\nit
Hence the orbit of a spinning particle moving in a Kerr-Newman space-time is
characterized by a new supersymmetry of the type (\ref{5.2}). Its physical
content is, that the projection of the spin in the direction obtained by
rotating the four velocity $\dot{x}^{\mu}$ by $f_{\mu}^{\:\:a}(x)$ is constant.

\section{N = 2 supersymmetry}

As another check on the previous results we can reproduce the well-known
conditions for the existence of a conventional $N = 2$ supersymmetry,
which is an independent supersymmetry generated by a charge $\tQ$ for which
the algebra closes on the hamiltonian:

\be
\left\{ Q, \tQ \right\}\, =\, 0, \hspace{3em}
\left\{ \tQ, \tQ \right\}\, =\, -2i H.
\label{8.1}
\ee

\nit
Since the bosonic constant of motion $Z$ now coincides with $H$, we have

\be
K^{\mu\nu}\, =\, g^{\mu\nu},
\label{8.2}
\ee

\nit
and therefore

\be
f^{\mu}_{\:\:a} f_{\nu}^{\:\:a}\, =\, \del_{\nu}^{\mu}.
\label{8.3}
\ee

\nit
The anti-commutativity of the two independend supercharges requires the
anti-symmetry of $f^{\mu\nu} = f^{\mu a} e^{\nu}_{\:\:a}$ as before.
As a result we can rewrite eq.(\ref{8.3}) in the form

\be
f^{\mu}_{\:\:\lb} f^{\lb}_{\:\:\nu}\, =\, - \del_{\nu}^{\mu}.
\label{8.5}
\ee

\nit
Moreover, since the covariant hamiltonian contains no explicit $\ps^4$-terms,
there is no $\ps^3$-term in $\tQ$:

\be
c_{abc} = 0, \hspace{2em} \Rightarrow \hspace{2em}
f^{\mu b} f_{\nu\:\: ;\mu}^{\:\:a}\, =\, f^{\mu b} f_{\nu\:\: ;\mu}^{\:\:a}.
\label{8.6}
\ee

\nit
Thus the existence of $N = 2$ supersymmetry requires an (almost) complex
structure $f^{\mu\nu}$, and this restricts the manifolds on which the models
are defined to be of K\"{a}hler type.

\section{Quantum theory}

Up until now we have presented the discussion of new supersymmetries in
the context of (pseudo) classical mechanics and geometry. The construction
can be carried over straightforwardly to the case of quantum mechanics,
by the usual replacement of phase-space co-ordinates by operators and
Poisson-Dirac brackets by (anti) commutators. In particular, we have

\be
\ba{lll}
\ps^a & \rightarrow & \dsp{ \sqrt{\frac{2}{\hbar}}\, \gam^a, }\\
 & & \\
\Pi_{\mu} & \rightarrow & \dsp{ -i \hbar D_{\mu}\, =\, -i \hbar \lh \pl_{\mu} -
\frac{1}{2} \og_{\mu ab} \sg^{ab} \rh. }
\ea
\label{9.1}
\ee

\nit
Here $\gam^a$ are the usual Dirac $\gam$-matrices, and $\sg^{ab}$ 1/4 times
their commutators. In terms of these operators the supercharge is replaced by
the Dirac operator

\be
\sqrt{\frac{2}{\hbar}}\, Q\:\: \rightarrow\:\: -i \hbar \gam \cdot D,
\label{9.2}
\ee

\nit
the dot denoting contraction with a vierbein, and the Hamiltonian becomes
a Laplacian, essentially the square of the Dirac operator.

The new supersymmetry charge is now replaced similarly by the operator

\be
\sqrt{\frac{2}{\hbar}}\, Q_f\:\: \rightarrow\:\: -i \hbar \gam^a \lh
       f^{\mu}_{\:\:a} D_{\mu} - \frac{1}{3!}\, c_{abc} \sg^{bc} \rh\,
       \equiv\, -i \hbar \gam \cdot F.
\label{9.3}
\ee

\nit
The independence of the symmetry transformations generated by these operators
on the spinor fields is expressed by the condition

\be
\left[ \gam \cdot D, \gam \cdot F \right]_+\, =\, 0.
\label{9.4}
\ee

\nit
The anti-commutator can be rewritten as a commutator by multiplication with
$\gam_5$. Therefore $\gam \cdot F$ corresponds to an operator which can be
diagonalized on solutions of the Dirac equation. This is the origin of the
separability of the Dirac equation in Kerr-Newman geometry observed in \ct{Ch}.
\nl\nl

\nit
{\bf Akcnowledgement} \nl
I am indebted to my collaborators Gary Gibbons of Cambridge University and
Rachel Rietdijk of Durham University for many pleasant and informative
discussions.

\end{document}